\documentclass[preprint,aps,showpacs]{revtex4}

\usepackage{graphicx}
\usepackage{dcolumn}
\usepackage{bm}
\newcommand{\be}{\begin{equation}}
\newcommand{\ee}{\end{equation}}
\newcommand{\bea}{\begin{eqnarray}}
\newcommand{\eea}{\end{eqnarray}}
\newcommand{\beb}{\begin{eqnarray*}}
\newcommand{\eeb}{\end{eqnarray*}}
\begin{document}
\preprint{LMPC/007}

\title{Exchange couplings in the magnetic molecular cluster Mn$_{12}$Ac}

\author{N. Regnault$^1$}
\email{Nicolas.Regnault@lpmc.ens.fr}
\author{Th. Jolic\oe ur$^1$}
\email{Thierry.Jolicoeur@lpmc.ens.fr}
\author{R. Sessoli$^2$}
\author{D. Gatteschi$^2$}
\author{M. Verdaguer$^3$}

\affiliation{ $^1$~Laboratoire de Physique de la Mati\`ere
Condens\'ee,\\ Ecole Normale Sup\'erieure, 24 rue Lhomond, 75231
Paris, France }

\affiliation{$^2$~Department of Chemistry, University of Florence,
Via della Lastruccia 3, 50019 Firenze, Italy}

\affiliation{$^3$~Laboratoire de Chimie Inorganique et Mat\'eriaux
Mol\'eculaires,\\ Universit\'e Pierre et Marie Curie, 4 place
Jussieu, 75252 Paris, France }

\date{April, 8, 2002}
\begin{abstract}
The magnetic properties of the molecular cluster Mn$_{12}$Ac are
due to the four Mn$^{3+}$ ions which have spins S=3/2 and the
eight Mn$^{4+}$ ions with spins S=2. These spins are coupled by an
exchange mechanism. We determine the four exchange couplings
assuming a Heisenberg-type interaction between the ions. We use
exact diagonalization of the spin Hamiltonian by a L\'anczos
algorithm and we adjust the couplings to reproduce the
magnetization curve of Mn$_{12}$Ac. We also impose the constraint
of reproducing a gap of $35$~K between a S=10 ground state and a
first excited state with S=9. We predict that there is an excited
level with S=8  at 37~K above the ground state, only slightly
above the S=9 excited state which lies at 35~K and the next
excited state is a S=9 multiplet at 67~K above the S=10 ground
state.
\end{abstract}
 \maketitle
\section{\label{sec:level1}Introduction}
Molecular nanomagnets consist of a few paramagnetic ions coupled
by exchange interactions and their properties are lying at the
borderline between quantum and classical behavior. The best
studied cluster~\cite{Lis} so far is certainly Mn$_{12}$Ac. At low
fields and low temperature its ground state can be described as a
single spin S=10. The reversal of this spin occurs via a
macroscopic relaxation time  at T=2~K. This
discovery~\cite{Sessoli93} has prompted a detailed examination of
its magnetic behavior because of the potential use for ultimate
information storage. If we concentrate only on magnetic ions, then
Mn$_{12}$Ac is made of an external ring of eight Mn$^{3+}$ ions
with spin S=2 and ring encloses a tetrahedron of four Mn$^{4+}$
ions with spin S=3/2. These spins are coupled by exchange so as to
lead to a S=10 ground state. This means that there is a
ferrimagnetic arrangement of the spins.  It is also known that the
S=10 manifold is split by anisotropy into sublevels with $-10\leq
S^z\leq +10$. The properties of this subset of levels have been
studied in detail by various experimental techniques because of
the possibility of macroscopic quantum tunneling of the spin. The
anisotropy responsible for the zero-field splitting of the S=10
manifold is smaller than the exchange interactions that determine
first of all the ferrimagnetic ground state structure. For example
the degenerate $S^z=\pm 10$ states are separated from the states
with $S^z=\pm 9$ (belonging to the {\it same} S=10 manifold) by a
gap determined by inelastic neutron scattering~\cite{MirebeauL99}
$\approx 14$~K. So a reasonable strategy to study magnetic
properties of this cluster is to ignore first any anisotropy terms
in a model spin Hamiltonian by use of the simplest Heisenberg
exchange and then, once these parameters are known, refine the
treatment by including higher-order anisotropy terms.

Recently the full magnetization curve M(H) of the cluster
Mn$_{12}$Ac has been obtained~\cite{explosive} by an experimental
technique using explosive compression of the magnetic flux to
access to very high fields in the megagauss range. Many if not all
transitions between levels with different total spin have been
measured. This kind of measurement is a direct probe of the inner
magnetic structure of the cluster i.e. of the higher-energy scale
couplings.

In this paper we determine the values of the exchange couplings
between the Mn ions by using a Heisenberg spin Hamiltonian to
reproduce the high-field magnetization data. We also add as a
constraint the value of the gap between S=10 and S=9. With these
values we are able to predict the energies and multiplicities of
some of the low-lying states above the S=10 ground state. This
explains some features of the existing inelastic neutron
scattering data that were not understood previously.
 In section II, we expose the details of the model Hamiltonian
and the numerical technique we have used. In section III, the
details of the fit are given. Finally section IV contains our
conclusions.

\section{The spin Hamiltonian}
\begin{figure}
\includegraphics[width=10cm]{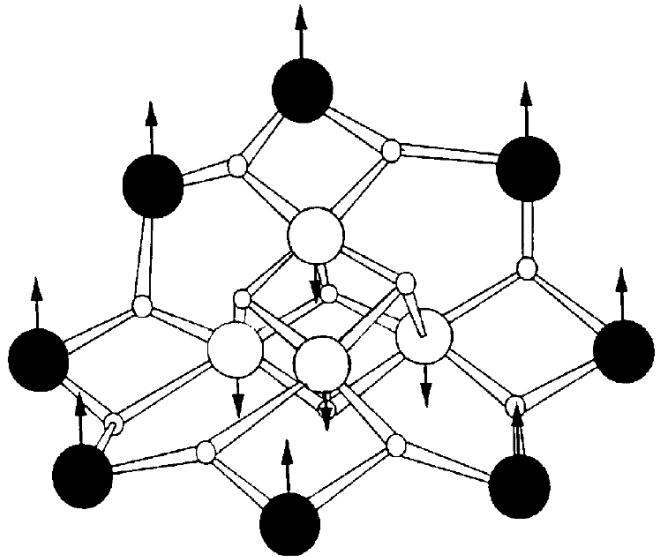}
\caption{\label{mn12mol} The core of the magnetic cluster
Mn$_{12}$Ac.}
\end{figure}

\begin{figure}
\includegraphics[width=10cm]{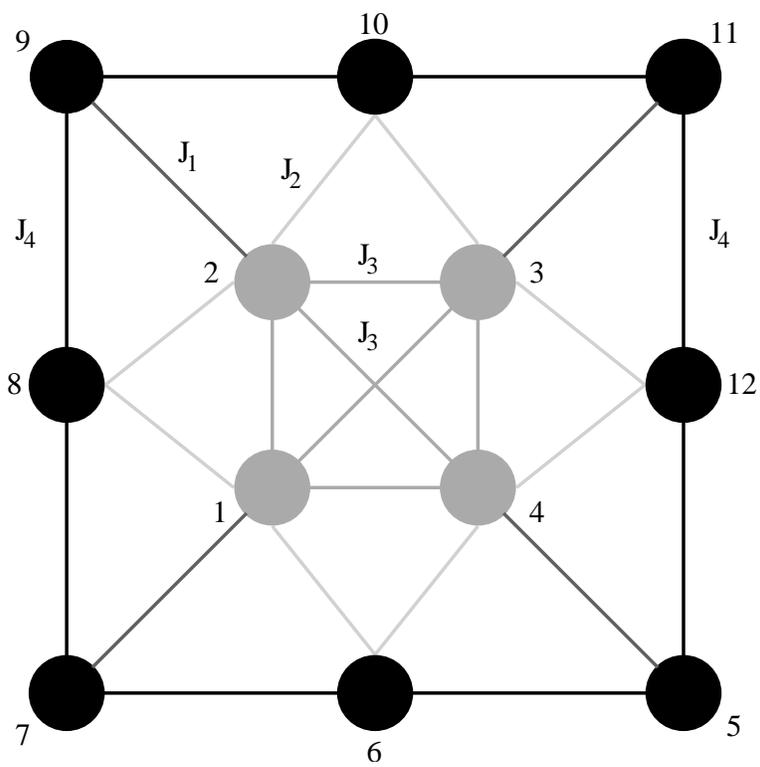}
\caption{\label{mn12spin} The spins in Mn$_{12}$Ac }
\end{figure}

The core of the magnetic cluster of formula
Mn$_{12}$O$_{12}$(CH$_3$COO)$_{16}$(H$_2$O)$_4$ is depicted in
Fig.~\ref{mn12mol}. There is an outer ring of eight ions Mn$^{3+}$
with spin S=2, $\mathbf{S}_{5},\dots \mathbf{S}_{12} $ that
encircles four ions Mn$^{4+}$ with S=3/2, $\mathbf{S}_{1},\dots
\mathbf{S}_{4}$. In Fig.~\ref{mn12spin} we draw a simplified
representation of the cluster taking into account only the
connectivity of exchange interactions~\cite{Caneschi91,Sessoli93}.
There are four most important exchange paths (bonds in the
following). They are J$_1$ which relate each spin 3/2 to a spin
S=2, there is also a J$_2$ coupling between the outer ring and the
inner tetrahedron involving four triangles i.e. eight bonds, then
J$_3$ between the four spins S=3/2 with a tetrahedral structure so
there are six such bonds, and J$_4$ between the S=2 ions which
means eight bonds. The simplest guess for the exchange is thus a
Heisenberg spin Hamiltonian~: \bea \mathcal{H} = &J_1& \{
\mathbf{S}_{1}\cdot\mathbf{S}_{7}+
\mathbf{S}_{2}\cdot\mathbf{S}_{9}+
\mathbf{S}_{3}\cdot\mathbf{S}_{11}+
\mathbf{S}_{4}\cdot\mathbf{S}_{5} \} \nonumber \\ +&J_2& \{
\mathbf{S}_{6}\cdot(\mathbf{S}_{1}+\mathbf{S}_{4})+
\mathbf{S}_{8}\cdot(\mathbf{S}_{1}+\mathbf{S}_{2})+
\mathbf{S}_{10}\cdot(\mathbf{S}_{2}+\mathbf{S}_{3})+
\mathbf{S}_{12}\cdot(\mathbf{S}_{3}+\mathbf{S}_{4})\} \nonumber
\\ +&J_3& \{ \mathbf{S}_{1}\cdot\mathbf{S}_{2}+
\mathbf{S}_{2}\cdot\mathbf{S}_{3}+
\mathbf{S}_{3}\cdot\mathbf{S}_{4}+
\mathbf{S}_{4}\cdot\mathbf{S}_{1}+
\mathbf{S}_{1}\cdot\mathbf{S}_{3}+
\mathbf{S}_{2}\cdot\mathbf{S}_{4} \} \nonumber \\ +&J_4& \{
\mathbf{S}_{5}\cdot\mathbf{S}_{6}+
\mathbf{S}_{6}\cdot\mathbf{S}_{7}+
\mathbf{S}_{7}\cdot\mathbf{S}_{8}+
\mathbf{S}_{8}\cdot\mathbf{S}_{9}+
\mathbf{S}_{9}\cdot\mathbf{S}_{10}+
\mathbf{S}_{10}\cdot\mathbf{S}_{11}+
\mathbf{S}_{11}\cdot\mathbf{S}_{12}+
\mathbf{S}_{12}\cdot\mathbf{S}_{5}\},\label{Ham}\eea where we use
the labeling of the spins given in Fig.~2. In this work we ignore
the effect of anisotropy and use the Hamiltonian Eq.(\ref{Ham}).
As a consequence the total spin is a good quantum number. The
total dimension of the Hilbert space of the magnetic degrees of
freedom is $10^8$ which is too huge for brute force
diagonalization. Previous
works~\cite{Sessoli93,Barrat97,Zvezdin96} have used various
approximate schemes usually based on some assumptions about the
relative order of magnitude of the J's. In the Florentine coupling
scheme~\cite{Sessoli93} one starts from a large antiferromagnetic
value for J$_1$. Hence the cluster is described as a first
approximation by four dimers Mn$^{3+}$-Mn$^{4+}$ with spin S=1/2
and four remaining Mn$^{3+}$ ions with spin S=2. It is then
feasible to treat the system of four spins S=1/2 and four spins
S=2. However it is not clear that this hypothesis is able to
explain the experimental results at hand. There are many results
from neutron scattering~\cite{MirebeauL99},
magnetization~\cite{Barbara98} and heat capacity~\cite{Gomes98}
that point to the presence of excited states not too far from the
S=10 ground state. It is thus desirable to obtain values of the
exchange parameters without recourse to any {\it a priori}
assumption of their relative strength. A recent study by exact
diagonalization~\cite{Raghu2001} using the point symmetry group of
the magnetic cluster concluded that J$_1$=215~K, J$_2$=85~K,
J$_3$=85~K, J$_4$=-64.5~K. Since these data refer to transitions
involving the lowest energy levels for a given spin value, the
L\'anczos algorithm appears to be particularly well suited.

The recent megagauss magnetization data~\cite{explosive} provide
further information on the strength of the exchange interactions.
We have used the L\'anczos algorithm~\cite{Golub} applied to the
Hamiltonian Eq.(\ref{Ham}). If we use conservation of spin
projection on a given axis, say {it z}, then the Hilbert space
dimensions are much reduced. The largest subspace corresponds to
$S^z=0$ with dimension 8581300, while the subspace in which we
have to find the ground state has dimension 817176 for $S^z=10$.
The corresponding dimensions are given in table~\ref{table1}.
Although the use of {\it total} spin would reduce further the
dimensionality, it is much more complicated to program in an
efficient way.

In a given subspace we use the iterative L\'anczos algorithm that
brings the Hamiltonian into tridiagonal form. This is done
typically in at most one hundred iterations which is enough to
obtain the ground state energy with a precision of $\approx
10^{-10}$. The advantage of the L\'anczos algorithm is that it
requires only to perform the product of the Hamiltonian on a
vector. Since spin Hamiltonians are very sparse matrices, one
needs only to store in fact two vectors to use the L\'anczos
algorithm. Even with the huge dimensions that appear in table~I,
this is feasible on present day computers.

If we now consider the effect of a magnetic
field~\cite{Golinelli}, then due to the rotational symmetry of the
Hamiltonian, the magnetic field couples to a conserved quantity
and we just have to shift energies~:
\be
E_{S^z}(B)=E_{S^z}(B=0)-g\mu_{\rm B}BS^z ,\label{fieldeffect} \ee
where $E_{S^z}(B=0)$ is an eigenenergy of (\ref{Ham}) in the
sector with spin projection $S^z$. As a function of the applied
field there will be crossings of levels with different values of
$S^z$. The magnetization curve of the cluster Mn$_{12}$Ac is a
series of discrete jumps and plateaus due to the finite size of
this spin system. It is only for the thermodynamic limit that one
gets smooth magnetization curves for spin systems. The critical
field corresponding to the transition $S^z$ to $S^z+1$ is given
by~:
\be
B_{crit}(S^z\rightarrow S^z+1) ={1\over g\mu_{\rm B} }
(E_{0,S^z+1}(B=0)-E_{0,S^z}(B=0)) ,\label{critfield} \ee where
$E_{0,S^z}(B=0) $ is the ground state energy in the sector with
given $S^z$. We have used g=2 in this paper.

\section{Reproducing Magnetization data}

To determine the exchange parameters, we have used the data coming
from various experimental sources. First it is well established
that the ground state has total spin S=10. These states are
exactly degenerate in our case since we make the simplifying
assumption of exact rotational symmetry. Then the gap between this
S=10 manifold and the first excited state with S=9 is known from
magnetic susceptibility measurements~\cite{Caneschi92} to be 35~K.
The remaining piece of high-energy information comes from the
megagauss experiment of ref.~\cite{explosive}. In this experiment
one measures the differential susceptibility $dM/dH$ vs. $H$. For
a system with a series of discrete jumps as Mn$_{12}$Ac, a spike
in the differential susceptibility corresponds to a change of the
spin of the ground state. The lowest-lying spike lies at $B_1=$
382~T and is interpreted~\cite{Zvezdin96} as the transition from
the ground state with S=10 to an S=11 state and then there are
three spikes at $B_2=$ 416~T, $B_3=$ 448~T, and $B_4=$ 475~T
corresponding to the crossing of S=12, 13, and 14 states
respectively. At higher fields there is a huge spike centered at
530~T which presumably corresponds to several unresolved
crossings, maybe S=15 and S=16. Above this field value it is
difficult to locate the other remaining transitions so no other
values were determined in Ref.~\cite{explosive}. We have used the
four values that are determined with good accuracy
$B_1,\dots,B_4$. We compute the corresponding theoretical values
$B_i^{\rm theo}$ and measure the quality of the fit by the
following quantity~:
\begin{equation}
\epsilon=\sum_{i=1}^{4}\left({B_i -B_i^{\rm theo}\over
B_i}\right)^2 . \label{chideux}
\end{equation}

\begin{figure}
\includegraphics[width=12cm]{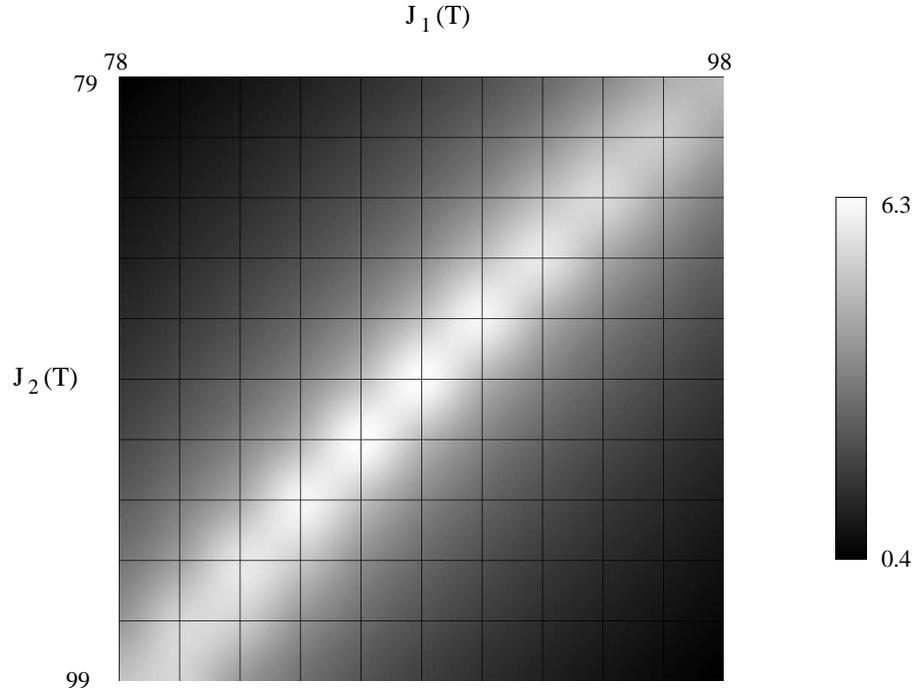}
\caption{\label{J1J2eps} Variation of the fitting parameter
$\epsilon$ as a function of J$_1$ and J$_2$. The remaining
couplings are taken to be J$_3$=-6.0~T (-8~K), J$_4$=17.0~T
(23~K). The grey levels are proportional to $-\log (\epsilon\times
10^{4})$.}
\end{figure}

\begin{figure}
\includegraphics[width=12cm]{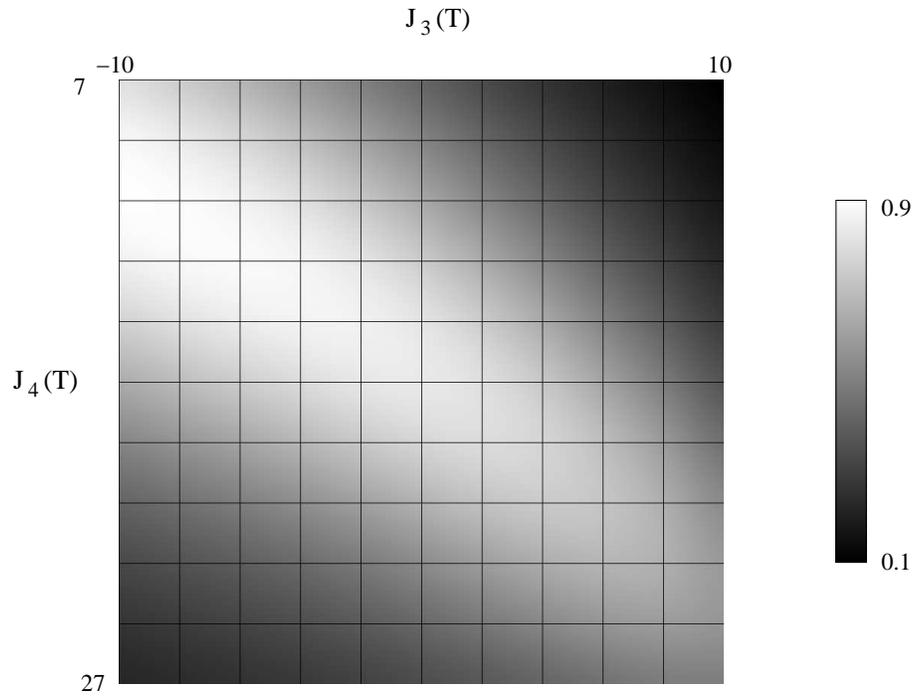}
\caption{\label{J3J4eps}  Variation of the fitting parameter
$\epsilon$ as a function of J$_3$ and J$_4$. The remaining
couplings are taken to be J$_1$=88.9~T (119~K), J$_2$=88.0~T
(118~K). The grey levels are now proportional to
$2/(1+\epsilon\times 10^{4})$.}
\end{figure}

We have performed calculations of the levels with S$^z$=9 up to
S$^z$=22 on a grid of values of the ratios of the exchange
couplings J$_2$/J$_1$, J$_3$/J$_1$ and J$_4$/J$_1$ between +2 and
-2 by steps of 0.1. In this range of parameters, we first reject
values for which S$^z$=10 is not the ground state. Then we look
for regions with small $\epsilon$ parameter defined in
Eq.(\ref{chideux}). In these regions, we fix the {\it absolute}
scale of energy via J$_1$ by requiring that the critical fields
should be equal to the experimental value $B_1=$ 382~T. This leads
to approximate values for the other critical fields
$B_2,\dots,B_4$. We next refine the search by including only the
regions in which the gap S=10-S=9 is close to its experimental
value of 35~K. This is now done by adjusting the {\it four}
dimensionfull couplings J$_1$,...,J$_4$. We observe that the
biggest effect on the overall spectrum structure is due to J$_1$
and J$_2$.   The best values are close to J$_1$ $\approx 119$~K,
J$_2$ $\approx 118$~K. This corresponds to $\epsilon\approx
10^{-4}$. In Fig.~\ref{J1J2eps} we plot the variation of the
fitting parameter $\epsilon$ in the J$_1$-J$_2$ plane close to the
best values of these parameters. If we search for a region of
small $\epsilon$ by tuning these two parameters only then we find
that this region is quite insensitive to the choice of J$_3$ and
J$_4$. This is illustrated in Fig.~\ref{J3J4eps} where $\epsilon$
is given in the J$_3$-J$_4$ plane.

\begin{figure}
\includegraphics[width=12cm]{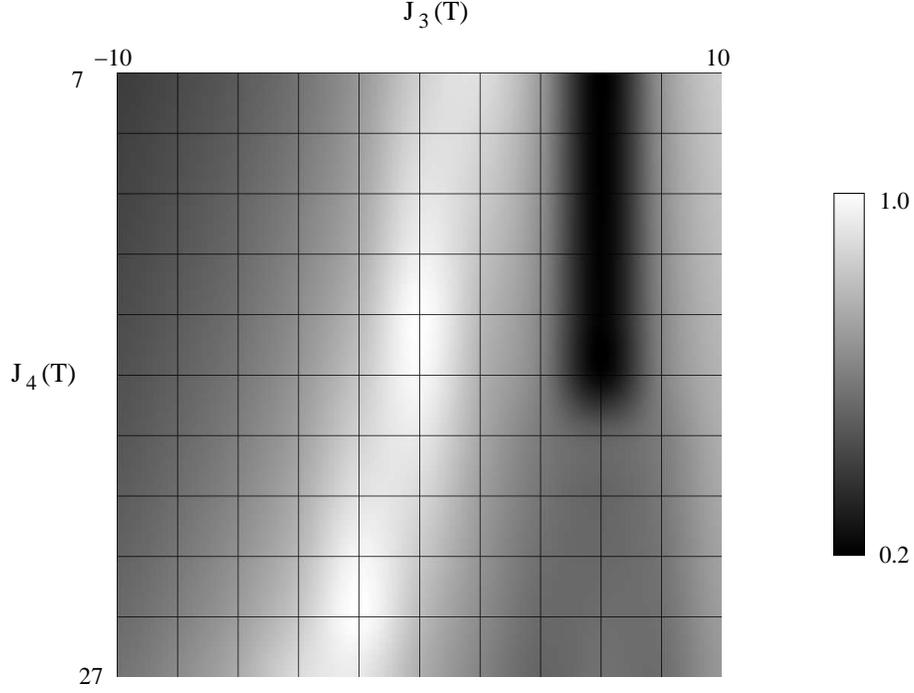}
\caption{\label{J3J4Gap} Variation of the gap as a function of
J$_3$ and J$_4$. The remaining couplings are taken to be
J$_1$=88.9~T, J$_2$=88.0~T. Grey levels are plotted via
$1/1+|(\Delta - 26)/26|$.}
\end{figure}

\begin{figure}
\includegraphics[width=12cm]{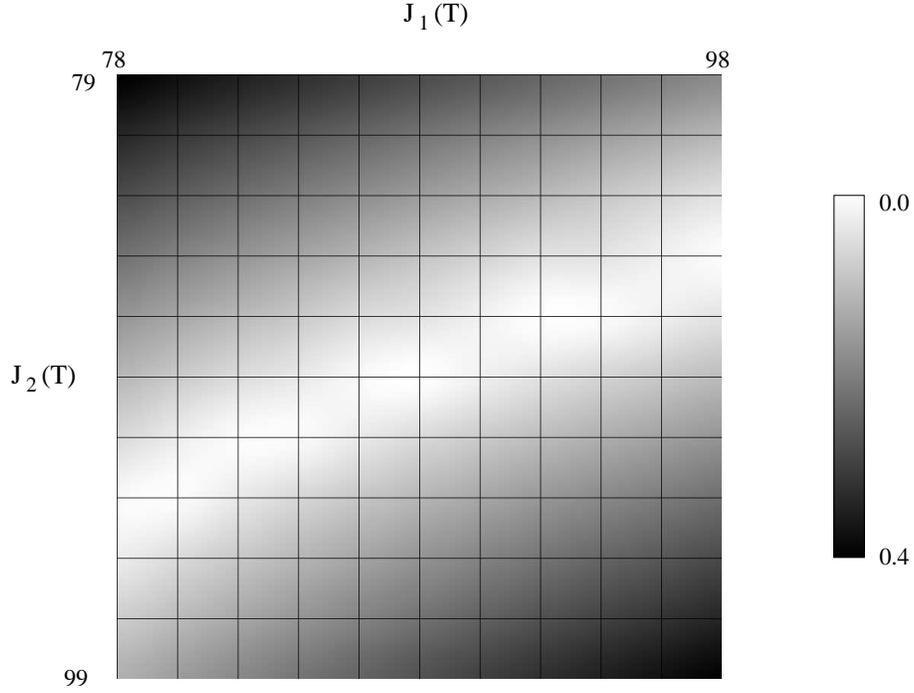}
\caption{\label{J1J2Gap} Variation of the gap as a function of
J$_1$ and J$_2$. The remaining couplings are taken to be
J$_3$=-6.0~T, J$_4$=17.0~T. Grey levels are given by the relative
variation $|(\Delta - 26)/26|$. }
\end{figure}

Our next observation is that once J$_1$ and J$_2$ are determined,
the remaining couplings J$_3$ and J$_4$ may be varied to obtain
more precisely the S=10-S=9 gap of 35~K. The agreement with the
gap is given in Fig.~\ref{J3J4Gap} where we have plotted the gap
value in the plane J$_3$-J$_4$. We find that J$_4$ is
significatively antiferromagnetic J$_4\approx 23$~K while it is
difficult to give a precise estimate for J$_3$~: it barely differs
significatively from zero. A tentative value is J$_3\approx -8$~K.
Taking into account the experimental uncertainties on the critical
fields, we estimate that J$_3$ and J$_4$ are determined with an
error bar of $\approx$ 6~K. The determination of J$_3$ and J$_4$
does not affect much the values of J$_1$ and J$_2$, as can be seen
in Fig.~\ref{J1J2Gap}.

As proposed in previous works, we find that J$_1$ is larger than
J$_3$ and J$_4$. However, we clearly need a second coupling J$_2$
which should be close to J$_1$ (taking into account the
uncertainties on all the experimental results). The part of the
spectrum relevant to magnetization data corresponding to this
preferred set of parameters is given in Table~II.

It is interesting to compare our results with those already
available in the literature. In the first works~\cite{Sessoli93},
the assumption of large J$_1$ combined with a perturbative
treatment lead to J$_1$=225~K,  J$_2$=90~K, J$_3$=90~K, J$_4$=0~K.
In fact this set of parameters does not lead to a S=10 ground
state, as first pointed out by Raghu et al.~\cite{Raghu2001}.
There is another set of parameters J$_1$=215~K,  J$_2$=85~K,
J$_3$=-85~K, J$_4$=-45~K suggested by
Chudnovsky~\cite{Chudnovsky96}. However this set which has a
correct S=10-S=9 ordering leads to a gap of 223~K, much too large.
A recent study by exact diagonalization~\cite{Raghu2001} concluded
that J$_1$=215~K, J$_2$=85~K, J$_3$=85~K, J$_4$=-64.5~K. This set
has a correct ordering of 9-10 levels with a gap which is adjusted
to the experimental value of 35~K but we find that it does not
lead to a satisfactory magnetization curve. This set gives
critical fields equal to $B_1=$ 192~T, $B_2=$ 239~T, $B_3=$ 356~T
and $B_4=$ 406~T (the $\epsilon$ parameter is 0.4 instead of our
value of $\approx 10^{-4}$).

With our preferred set of exchange parameters we can then compute
some of the excited levels that are above the first excited
manifold with S=9. We find that there is a S=8 manifold sitting at
37~K above the S=10 ground state so very close to the S=9 states
that have been used to constrain our fit. Then there is a S=9
multiplet which is found at 67~K above the ground state. This
picture is close to what is found from susceptibility
measurements~\cite{Caneschi92} where two close S=9 levels are
necessary to reproduce the low-temperature behavior. However the
scheme is not exactly the same~: the multiplicities do not
coincide.

Our new picture has some interesting consequences for the
interpretation of neutron data. There is some evidence from
inelastic neutron scattering~\cite{Hennion97} for a mode at 1.2
THz i.e. 70~K, this may correspond to the second S=9 multiplet
that we find at 67~K. This is allowed by neutron scattering
selection rule starting from the S=10 ground state. It is then
normal that neutrons do not see the S=8 states at 37~K = 0.73 THz
because of the selection rules however it not yet clear why
neutrons do not see the first excited state with S=9 at 35~K =
0.72 THz.

\section{Conclusions}

We have studied the energy levels of a Heisenberg spin model
Hamiltonian appropriate to describe the magnetic cluster
Mn$_{12}$Ac. We obtain a determination of four exchange
couplings~: J$_1$=119~K,  J$_2$=118~K, J$_3$=-8~K, J$_4$=23~K.
Such a set of parameters reproduces the magnetization curve
observed in megagauss experiments, leads to a S=10 ground state
and a gap of 35~K to a first excited level with S=9 as measured
experimentally. The numerical method we have used does not rely
upon any approximations, so the main source of uncertainty in the
values we quote comes from the measurement of the critical fields
in the magnetization process. We predict that there is an excited
level with S=8  at 37~K only slightly above the S=9 excited state
which lies at 35~K and the next excited state is a S=9 multiplet
at 67~K above the S=10 ground state. These findings explain part
of the existing neutron scattering data.

\begin{acknowledgments}

We wish to acknowledge support from the European community
(Molnanomag HPRN-CT 1999 0012), the French Ministry of Research
and the Region Ile-de-France (Chaire Blaise Pascal 2001). We thank
also Andreas Honecker and Bruce Normand for a careful reading of
the manuscript.

\end{acknowledgments}

\newpage

\begin{table}
\caption{\label{table1} Hilbert space dimension of subspaces of
given S$^z$ for the system of eight spin-2 and four spin-3/2.}
\begin{ruledtabular}
\begin{tabular}{lcr}
$S^z$&Dimension\\ \hline 22 & 1\\ 21 & 12\\ 20 & 78\\ 19  &
364\\18& 1361\\ 17&4312\\ 16&11968\\ 15&29744\\ 14&67216\\
13&139672\\ 12&269148\\ 11& 484144\\10& 817176\\9& 1299632\\8&
1954108\\7& 2785384\\6& 3772176\\5& 4862352\\4& 5974048\\3 &
7003944\\2& 7842070\\1& 8390440\\0&8581300\\

\end{tabular}
\end{ruledtabular}
\end{table}

\begin{table}
\caption{\label{table2}Ground state energies in each fixed $S^z$
sector with corresponding critical magnetic fields for coupling
constant set J$_1$=88.9~T ($\approx$ 119~K), J$_2$=88.0~T
($\approx$ 118~K), J$_3$=-6.0~T ($\approx$ -8~K) and J$_4$=17.0~T
($\approx$ 23~K) for which the gap is 26~T ($\approx$ 35~K).}
\begin{ruledtabular}
\begin{tabular}{lcr}
S$^z$ & Energy (T) & Critical fields (T)\\ \hline 10&-3196.2&\\
           &&381.0\\
11&-2815.2&\\
           &&411.7\\
12& -2403.5&\\
           &&444.1\\
13& -1959.4&\\
           &&477.6\\
14& -1481.8&\\
           &&512.5\\
15 &-969.3&\\
           &&548.4\\
16 &-420.9&\\
           &&584.9\\
17 &164.0&\\
           &&622.1\\
18 &786.1&\\
           &&659.7\\
19 &1445.8&\\
           &&697.2\\
20 &2143.0&\\
           &&733.9\\
21 &2876.9&\\
           &&764.9\\
22& 3641.8&
\end{tabular}
\end{ruledtabular}
\end{table}

\newpage

\end{document}